*Article*

# Functional Model of Residential Consumption Elasticity under Dynamic Tariffs

**Kamalanathan Ganesan [1,2]\*, João Tomé Saraiva [1,2] and Ricardo J. Bessa [2]**

1. Faculty of Engineering, University of Porto (FEUP), 4200-465 Porto, Portugal; jsaraiva@fe.up.pt
2. INESC Technology and Science (INESC TEC), 4200-465 Porto, Portugal; ricardo.j.bessa@inesctec.pt, kamalanathan.ganesan@inesctec.pt
\* Correspondence: kamalanathan.ganesan@fe.up.pt

**Abstract:** One of the major barriers for the retailers is to understand the consumption elasticity they can expect from their contracted demand response (DR) clients. The current trend of DR products provided by retailers are not consumer-specific, which poses additional barriers for the active engagement of consumers in these programs. The elasticity of consumers' demand behavior varies from individual to individual. The utility will benefit from knowing more accurately how changes in its prices will modify the consumption pattern of its clients. This work proposes a functional model for the consumption elasticity of the DR contracted consumers. The model aims to determine the load adjustment the DR consumers can provide to the retailers or utilities for different price levels. The proposed model uses a Bayesian probabilistic approach to identify the actual load adjustment an individual contracted client can provide for different price levels it can experience. The developed framework provides the retailers or utilities with a tool to obtain crucial information on how an individual consumer will respond to different price levels. This approach is able to quantify the likelihood with which the consumer reacts to a DR signal and identify the actual load adjustment an individual contracted DR client provides for different price levels they can experience. This information can be used to maximize the control and reliability of the services the retailer or utility can offer to the System Operators.

## 1. Introduction

*1.1 Context and Motivation*

The current residential demand response (DR) programs' engagement is far from its achievable potential because of a lack of addressing the barriers that largely vary between contracted individual consumers [1]. The aggregators and retailers must incorporate possible directions to modify consumption patterns along with their DR signals to clients, who are allowed to take free actions in reaction to the signals (e.g., economic, environmental) they receive. This will lead to confident participation of the consumers without their current concerns of energy security and privacy. This approach, on the other hand, will cause some concerns for the retailers regarding the reliability of the bids they make in the wholesale market.

Engaging the residential consumers and providing the best tariffs for their randomized behaviors is one of the major barriers identified in DR studies [2]. One of the major assumptions made by competitive electricity markets around the world which envision implementing (or currently work with) DR programs is that they assume the end-users (specifically residential consumers) will adopt the best decision available to them [3]. This generally results to assuming they react to DR prices as expected, accept direct load control and home automation, and participate in planned and predicted household activities that facilitate response to the price signals sent to them. But this is not the case from a behavioral economics perspective as consumers have limited knowledge of the benefits from actively participating in these programs. The actors' behaviors are more inclined to changes, which will not always be predictably proportional to prices [2, 4].

The most common residential DR programs involve customers shifting load manually in response to price signals. However, research has also looked at home automation systems that monitor and control the consumption of typical household appliances, ranging from AC's, washing machines, dishwashers to electric water heaters. From a general perspective, even with advanced metering technologies and reasonable experience with DR programs, the adoption of these strategies at the residential level is still limited. This could be due to the reason that DR pricing schemes could eventually lead to rebound effects, such cases when day-ahead hourly prices are applied new peak demand periods may arise [5]. However, it is reported that a considerable amount of residential load will be available for shifting for the utilities to balance the system. Nevertheless, the investments made by households on smart devices will not prove to be economic from their perspective.

DR resources cannot yet completely compete on an equal level with traditional (generation) resources. In addition, for household consumers, it is not easy to offer services on an individual basis. Therefore, DR service providers could for instance aggregate demand response of multiple similar end-consumers and sell it to the market as an aggregated bundle consumer [6-8]. In this case, service providers can make a pool of combined loads that they sell as a single resource and as such help individual consumers to value their flexibility potential.

*1.2 Related Work and Contributions*

One main challenge while studying a DR program is the ability of the model to distinguish a normal consumption from a DR instigated consumption. Lopes et al. [9] and Navigant [10] used a non-regression technique which compares the mean hourly consumption on event days to that of non-event days for the same household. George et al. [11] and Ericson [12] used a multiple regression technique where the regression coefficients are the estimates of the program effects and the events are the independent variables. Herter et al. [13] compares CPP households' average peak-time weekday loads to their normal-time average weekday loads for the same temperature (to avoid temperature bias) to identify the change in demand. Zhou et al. [14] used machine learning techniques to predict residential electricity consumption with non-parametric hypothesis to identify the reduction in household consumption during DR events. However, all these studies fail to quantify the likelihood with which the consumers alter their consumption behavior during DR events.

In [15], the author's idea was also to model an adaptive learning agent with reinforcement learning. A profit-seeking intelligent retailer makes repeated interactions with its clients to adapt its DR strategy based on the price elasticity of demand. The consumers have assumed/allotted self-elasticity values which determine their demand reduction potential during DR events. The drawback with using reinforcement learning in understanding consumer behavior is that one cannot establish a causal relationship linking the current state of the environment to the previous action. This could lead to undesirable action selection policy for the next state, and thus deviate from the actual behavior of a consumer. This is one of the reasons why both these studies have resorted to using clustered consumers' load and elasticity profile instead of individual consumers.

In a theoretical study done by H. C. Gils [16], substantial potential for demand response was found amongst residential consumers. According to some sources, *"It is now clear to policymakers that Europe will not be able to achieve its energy policy goals in a secure and cost-efficient manner unless the energy system becomes more flexible. DR and consumer empowerment are understood as integral parts of the Energy Union and the Clean Energy Package for all Europeans because they help to reach a competitive, secure, and sustainable economy"* [17]. Europe is now expanding on this model by giving individuals and communities the right to produce, store and sell energy. The revised Renewable Energy Directive RED II [18] and the Internal Electricity Market (IEM) Directive [19] introduced the concepts of Renewable Energy Community (REC) and of Citizen Energy Community (CEC), as part of the final Clean Energy Package. As such, Europe is giving consumers a way to organize themselves individually, or as a community, to ensure a more open energy-sector while empowering consumers in the context of the energy markets.

One research work that closely resembles one of the goals of this present paper was done by authors in [20] who relied on modelling DR as a reinforcement learning problem and cluster consumers based on their ability to provide curtailment during DR events. This process is done by an intelligent retailer agent, who collects data from their contracted clients and clusters them based on behavior properties. The main focus in this study was to identify a consumer's shiftable appliance activity and the population that are necessary in making real-time demand commitments.

To identify suitable users for DR, several studies, such as [21-27] have relied upon using historical household consumption data and categorizing them based on different load profiles with clustering techniques such as K-means clustering [23, 24, 28], fuzzy K-means [22], Fast Search and Find of Density Peaks (FSFDP) [25], support vector clustering [29], subspace projection based clustering [30], hierarchical clustering [26], expectation maximization algorithm [21], self-organizing maps [27], similarity measure between pairs of customers using Dynamic Time Wrapping (DTW) [31], clustering based on Artificial Neural Networks (ANN) [32], and etc. The goal of these studies was to evaluate the variability of consumption behavior for each cluster to quantify the household clusters' potential of price-based demand response.

Other studies such as [13, 33] carry out experiments with a control group (households who have a normal/default price-tariff) and a treatment group (households who experience new pricing-tariffs). In [33] it is used household energy consumption and income levels to cluster consumers, and [34] relies on house types and climatic zones as features to cluster clients and study their willingness to participate in DR events. They identify suitable DR consumers by comparing DR participants' mean critical weekday loads of the treatment group to the average normal weekday loads of the control group for the same period.

Most of the work reported in the literature deals with flexibility based on the potential decrease in residential demand a DR program can realize from participating households. However, potential increase in demand is also a sort-out characteristic in residential DR that is often not expressed or studied. The studies in the literature also assume user compliance while clustering (segmenting) households which might not be the best representation of their consumption behavior. For example, households with periodic and consistent consumption pattern will be less willing to alter their loads [14]. Therefore, it is necessary to address the causality of behavior patterns amongst residential DR consumers.

It is imperative that the potential of residential DR should not be overestimated in models because of unrealistic assumptions (such as flexibility based on baseline load consumption) about consumer engagement. According to these ideas, this paper aims to provide guidelines that could prove vital in setting up DR programs that will overcome the current market barriers and increase the adoption of such programs that at the end will/should eventually result in novel engagement strategies. In many cases, experiments are not possible either because of the lack of available funds for such vast consumer experiments, or not enough trust or motivation from the consumers to engage in pilot projects or it is just impossible/unethical to collect. Another reason to explain this difficulty is that many consumers who participate in such projects might be very similar, and we do not want to unfairly exclude any of them from a core customer belief (such as high DR price leads to more flexibility). In such situation, our proposed technique is to use observational methods aiming at understanding the causal effects in the absence of an experiment. This type of technique is frequently used in medical studies, where such types of observational research is implemented. These techniques are typically cheap and are generated as a side effect of normal operation. A more detailed explanation of our Causality Framework is available at [35, 36] and is briefly presented in Appendix A. The objective of this framework was to identify the causal relationship between the DR price and consumption. The methodology uses causal inference theory to estimate the elasticity of domestic electricity consumption in response to dynamic time-of-use tariff. The convenience associated to the application of this causal model is to use the available information regarding electricity consumption from before and after a DR event instead of two different (experimental and control) groups. A Robin's g-method based on parametric g-formula and kernel regression is used as a causal effect estimator and a Bayesian structural time-series model is used to determine the extent to which the price changes impact on the consumer's consumption. This gives us a likelihood probability of the price having a causal effect on the consumption. The model provides the potential to help the utilities or retailers to estimate the available elasticity from each of their contracted consumers in the scope of DR programs for each of the price signals they have experienced during the DR program.

Thus, in this paper, using the elasticity values we have from the causality inference framework, we build a functional model for the consumer's elasticity. This functional model will output a curve relating the load adjustment (elasticity provided) to the prices experienced by the consumers under different scenarios (outside temperature, time of day, day, month, etc.). This will eventually provide us with an elasticity behavior model for each consumer. The upside of using this technique is that it incorporates useful information, and the considered parameters will relate to the absolute effects that can occur in electricity consumption changes.

The main original contributions from this paper can be classified into two folds. The first one is a consumer probability score model based on the Dirichlet-multinomial distribution and Bayesian inference. The model is built based on the observed elasticities and their weights from the causality framework. With model parameters such as the beta-distribution and pseudo-counts relating to each consumer, a Bayesian inference is used to identify the distribution of a consumer's acceptance probability (when a set of consumers are grouped together for an event). The second original contribution is an approach using Bayesian probabilistic programming to use the elasticity of the consumer's demand response to the offers they receive from the market/retailers and use them to estimate probabilities of elasticity values for unseen (unexperienced) tariffs. This will give a clear picture to the retailers in terms of the impact such offers make on a specific client, i.e. if the tariff is lower or increases twice its rate for a given period, how would that client react to such variations. It gives an insight into what price levels each consumer was more susceptible to and what was the untapped potential from underperforming consumers.

*1.3 Structure of the Paper*

After this introductory section, Section 2 describes the residential consumption data from London electricity pricing trial (available in open access) and general methodology, Section 3 describes the developed Bayesian probabilistic methodology for estimating consumers probability score for experienced DR prices and Section 4 describes the estimation of consumption elasticity to unexperienced DR prices, and Section 5 enumerates the most relevant conclusions. Finally, Appendix A provides a brief description of the adopted Causality model.

**2. Data Collection and Methodology**

*2.1 Dataset Description*

The Low Carbon London (LCL) project was UK's first residential sector, time-of-use electricity pricing trial. UK Power Networks and EDF Energy jointly performed this project. The trial involved 5667 households organized in two groups, one the target group and the other the control group. The group that was influenced by dynamic Time-Of-Use tariffs (dToU group) consisted of 1122 households and the control group, which remained with existing non-dynamic Time-Of-Use tariff (non-ToU), consisted of 4545 consumers [37]. The electricity consumption was measured every 30 mins from 2012 (July–December) till 2013 (full year) for the dynamic Time-Of-Use (dToU) group from which we selected 81 clients. The dataset also contained tariff information, which comprised of three rates for the year 2013: Default is 0.1176 £/kWh; high is 0.6720 £/kWh; low is 0.0390 £/kWh. These rates were applicable to the dToU group only for the year 2013, whereas a flat tariff of 0.1176 £/kWh was maintained for the year 2012. Customers were informed of upcoming price changes one day ahead of delivery via notifications that appeared on their smart-meter linked in-home-display and also, if requested, via short message service (SMS) messages to their mobile phones. The price events were based on system balancing (lasted between 3 h to 12 h) and distribution network constraint management signals (which lasted for nearly 24 h) received by the retailers.

We use 81 dToU clients who have the best-recorded values and after data cleaning provide the best featured collection of consumers. The LCL dataset included an appliance survey dataset which collected information such as appliance ownership, physical parameters of the premises (either, house, apartment or mobile house) and basic details of its occupants such as energy decision maker, household size and age). In our analysis, all 81 clients were considered from the "house" category, and with household size "2 and higher" to have a consistent result. Other demographic information was collected in their survey but was not publicly available for using in our analysis. Since weather related information was not provided with this dataset, daily as well as half-hourly weather data for the years 2012 and 2013 were collected from an online source [38]. The new dataset was designed to include the following features for each client:

a) average power (kW) per 30 min interval;
b) hour of consumption;
c) price associated with the consumption period;
d) atmospheric temperature, humidity, pressure, visibility, wind direction, wind speed and weather conditions (example. Cloudy, clear, rainy fair, etc.);
e) current time of an event, minute, month, day of the week, week of the year, weekday number;

Building electricity consumption does have a direct causal relationship with the building occupancy rate. Specific loads such as, plug-loads have a stronger correlation with the occupancy rates as shown by Kim et al. [39]. Their study estimates about 10%-40% of energy can be saved by factoring occupancy rates into building energy consumption. As our dataset, does not include the occupancy rate, we rely on differentiating the "daily routine" and "occupancy lifestyle" by the day of the week feature in our Causality Inference model, that estimates our elasticities used in the Elasticity Behavior model.

*2.2    Methodology - Bayesian Probabilistic Programing Framework*

2.2.1 General Modelling Aspects

The adopted approach to tackle this problem adopts a data-driven Bayesian inference with probabilistic models. We need to include uncertainty in our estimate considering the data we have collected about each consumer. Second, we also must incorporate prior beliefs about the situation into this estimate. If we can assume the temperature variations during days & seasons, occupancy rate during weekday & weekends, probability of acting towards DR signal, etc., then surely this should play some role in our estimate. Thus, to allow our model to express such uncertainty and incorporate prior information into the estimates, we rely on Bayesian inference techniques. This section explores the problem of estimating probabilities from our data in a Bayesian Framework, along the way learning about probability distributions, Bayesian Inference and probabilistic programming.

Bayesian Inference is based in the fact that the truth is somewhere in-between the prior knowledge of the scenario under consideration, and the observed data. Building a constructive relationship between the prior belief and the observed data can result in the posterior belief of a model.   This is expressed using the Bayes Formula given by (1).

$$p(\theta|X) = \frac{p(X|\theta) \cdot p(\theta)}{p(X)} \tag{1}$$

$$p(X) = \sum_{\theta} p(X|\theta) \cdot p(\theta) \tag{2}$$

$$p(\theta|X) = p(X|\theta) \cdot p(\theta) \tag{3}$$

In (1), X is the data variable whose value we observe as x and estimate p(θ), a parameter of an unknown variable Y. We see that the Bayes Theorem expresses the posterior density p(θ|X), in terms

of likelihood p(X|θ) and prior p(θ), along with the marginal likelihood p(X) which is usually the normalization term given by Equation (2). We then can express the posterior probability distribution function p(θ|X) as described by Equation (3).

The Markov Chain Monte Carlo (MCMC) method can be used for estimating the parameters (θ) of a random model (such as Equation (1)), by running the model for a large number of iterations within a reasonable amount of time. The model's likelihood function and prior distribution will be used to draw the trajectory (such as trees) through the space for all possible values of θ. So, in a way, one should make sure the prior and the likelihood relationship to be as accurate as possible to estimate the best estimate fit for θ based on the observations. The posterior density given by equation (2) is used to guide the Markov chain $S = (θ_1,..θ_n)$ through the parameter space for all the possible value settings of θ. After drawing many samples from the defined model, we converge to an approximately desired posterior along with their uncertainties. Probabilistic Programing (PP) is a programming technique to unify probability models and inferences for these models to help make better decisions when faced with uncertain situations. The idea behind PP is to bring the inference algorithms and theory from statistics combined with formal semantics, compilers, and other tools to build efficient inference evaluators for models and applications from Machine Learning.

In other words, probabilistic programming is a tool for Bayesian statistical modeling. Probabilistic thinking is an incredibly valuable tool for decision making. From economists to poker players, people that can think in terms of probabilities tend to make better decisions when faced with uncertain situations. Probabilistic programming (PP) allows flexible specification of Bayesian statistical models in code. In our work, we use a Python framework called PYMC which performs the probabilistic programming using Theano, that allows for automatic Bayesian inference on user-defined probabilistic models. Model assumptions are recorded with prior distributions over the variables that influence the model. At the execution level, the model will launch the inference procedure to automatically compute the posterior distribution of the parameters of the model based on observed data. This is done in a way that the inference adjusts the prior distribution using the observed data as a base, to give a more precise mode. Recent advances in Markov Chain Monte Carlo (MCMC) sampling allow inference on increasingly complex models. PYMC features next-generation Markov Chain Monte Carlo (MCMC) sampling algorithms such as the No-U-Turn Sampler (NUTS, [40]), a self-tuning variant of Hamiltonian Monte Carlo (HMC, [41]). HMC and NUTS take advantage of gradient information from the likelihood to achieve much faster convergence than traditional sampling methods, especially for larger models. NUTS also has several self-tuning strategies for adaptively setting the tunable parameters of Hamiltonian Monte Carlo.

## 3. Consumer Response Probability Score

The first main original contribution from this paper is based on a methodology using the Dirichlet-multinomial distribution [42]. In Bayesian statistics, the parameter vector for a multinomial function is drawn from a Dirichlet Distribution, which forms the prior distribution for the parameter. In Bayesian probability theory, conjugate distribution or conjugate pair means a pair of a sampling distribution and a prior distribution for which the resulting posterior distribution belongs into the same parametric family of distributions than the prior distribution, and the prior is called a conjugate prior for this sampling distribution. Even with increasingly better computational tools, such as MCMC, models based on conjugate distributions are advantageous. One such conjugate distribution is the Beta-Binomial distribution which is based on binary choices, and a generalization of Beta distribution is called the Dirichlet-Distribution which is based on multiple choices. It corresponds to a family of discrete multivariate probability distributions on a finite support of non-negative integers. The Dirichlet distribution models the probabilities of multiple mutually exclusive choices, parameterized by "alpha" (which is referred to as the concentration parameter) and represents the weights for each choice. In the literature, we encounter the Dirichlet distribution often in the context of topic modeling in natural language processing, where it is commonly used as part of a Latent Dirichlet Allocation (or LDA) model. However, for our purposes, we look at the Dirichlet-Multinomial in the context of acceptance probability for the responses to the DR price signals received by the consumers.

*3.1 Methodology*

To identify the probability of accepting/acting to change consumption for a given price signal, we use Bayesian methods to look for the posterior probability. Thus, we construct a model for this situation with the series of observed elasticities (as collected from the Causality Framework developed in [36]) and their ranking (model defined weights). For our model, we have a case where unlike in binomial distributions (just having 2 possible outcomes) we have many outcomes which means that we are dealing with a multinomial distribution. Our multinomial distribution is characterized by the following parameters;

| | | |
|---|---|---|
| $n$ | = | sum of the weights associated with elasticity |
| theta ($\theta$) | = | Dirichlet distribution of alpha (vector of probabilities for each of the outcomes) |
| alpha ($\alpha$) | = | concentration parameter |
| $p$ | = | event probability for each consumer (parameters of multinomial) |
| $c$ | = | observed elasticity weights data |
| $h$ | = | Time period (hour) of the observed data |
| $i$ | = | Either "Price" or "Consumer" depending on model chosen |
| $k$ | = | number of outcomes (either number of prices or consumers) depending |

Our goal is to estimate the posterior distribution for the probability of observing each consumer $p_k$, conditioned on the data and hyperparameters $\alpha\_k$, for each hour-$h$. Our model comprises of a multinomial model with a Dirichlet distribution as the prior and a specified hyperparameter vector as described by Equation (4).

$$\text{solve for (Multinomial-model)} \quad (p \mid c, \alpha) \quad (4)$$

$$\text{where,} \quad \alpha_i > 0 \quad \text{with } (\alpha_1, \ldots \alpha_k)_h$$

$$\sum c_{i,h}^k = n_h \text{ and } \sum_k \theta_k = 1$$

| | | | | | |
|---|---|---|---|---|---|
| $p \mid \alpha$ | = | = | $(p_1, \ldots, p_k)_h$ | ~ | $(k, \alpha)$ |
| $c \mid p$ | = | = | $(c_1, \ldots, c_k)_h$ | ~ | $(k, p)$ |
| theta ($\theta$) | ~ | ~ | Dirichlet ($\alpha$) | | multinomial parameter |

Our distribution (4) is parameterized by the vector alpha which is called a hyperparameter because it is a parameter of the prior and it has the same number of elements as k as our multinomial parameter c. The assumptions are that each observation (elasticity rank for each hour and for each consumer) corresponds to an independent trial, and our initial prior is that each trial is equally represented (i.e., all trials have an equal chance of occurring).

We have 81 discrete choices from the LCL dataset (corresponding to the 81 consumers) each of them having an unknown probability with 24 total observations (24 elasticity weighted values for each consumer). Our objective is to identify the theta Dirichlet distribution for the given price and the hour based on the elasticity weights identified under prior conditions (causality framework). It should be noted that our initial alpha is set as a pseudo-count. This means that it dictates the effectiveness or reliability of our consumer. It is an amount added to the number of observed responses in order to change the anticipated probability in our model of those data. Larger pseudo counts will have a greater effect on the posterior estimate while smaller values will have a smaller effect and will let the data dominate the posterior. If we are confident about a particular consumer's performance, we can increase or decrease this parameter to take effect in the model.

We set up a model to estimate the posterior probability for theta (i.e., the probability of responses for each price for each hour) for different consumers. We will use the Dirichlet distribution variable from PYMC model for this purpose along with our observed variable (consumer elasticity ranking and unobserved (theta) variables. For the prior on theta, we will assume a non-informative uniform distribution, by initializing the Dirichlet prior with a series of 1s for the parameter alpha, one for each of the k possible outcomes. This shows that for every hour, the possibility of accepting a price is always available as our model is reduced to a 24 h scale for the elastic behavior. The expected value of a multinomial for each hour with Dirichlet priors can be expressed as given by (5).

$$E[p \mid c_h, \alpha] = \frac{c_i + \alpha_i}{n + \sum_k \alpha_k} \tag{5}$$

And the Dirichlet probability density function is expressed as (5.6).

$$p(\theta_1, \ldots \theta_k \mid \alpha_1, \ldots \alpha_k) = \frac{1}{B(\alpha)} \prod_{i=1}^{k} \theta_i^{\alpha_i - 1} \tag{6}$$

In this expression, $B(\alpha)$ is the multinomial beta function given by (5.7) which normalizes the distribution (alpha) making sure that the integral equals 1.

$$B(\alpha) = \frac{\prod_{i=1}^{k} \Gamma(\alpha_i)}{\Gamma(\sum_{i=1}^{k} \alpha_i)} \tag{7}$$

The posterior distribution given by Equation (8a) is solved using Bayesian inference, where we build the model and then use it to sample (to identify discrete outcomes) from the posterior to approximate the posterior with No-U-Turn Sampler method in PyMC3. We calculate the posterior probability of "c" by integrating out theta ($\theta$) as indicated from (8a) to (8d).

$$P(c|\alpha) = \int P(c,\theta|\alpha) \, d\theta = \int P(c|\theta) \, P(\theta|\alpha) \, d\theta \tag{8a}$$

$$= \int \left( \prod_{i=1}^{k} \theta_i^{N_i} \right) \left( \frac{1}{B(\alpha)} \prod_{i=1}^{k} \theta_i^{\alpha_i - 1} \right) d\theta \tag{8b}$$

$$= \frac{1}{B(\alpha)} \int \prod_{i=1}^{k} \theta_i^{N_i + \alpha_i - 1} \, d\theta \tag{8c}$$

$$= \frac{B(N + \alpha)}{B(\alpha)} \tag{8d}$$

The result is not just one number but rather a range of samples that lets us quantify our uncertainty. Our model draws 5000 samples from the posterior in parallel computing (5000 * 4 for a 4 core machine). We adjust the model to use the 1st 1000 samples as tuning samples, and the inferences are made only after the 1st 1000 samples.

This results in trace samples as shown in Figure 1, which we use to estimate the posterior probability distribution. The estimate p is the kernel density estimate for the sampled parameters (PDF of the event probabilities) along with uncertainty. The point estimate of p is the mean of the posterior trace samples and the Bayesian equivalent confidence interval (called credible interval) are the 95% and 5% highest probability density. The uncertainty in our posterior reduces with a greater number of observations. We have the different elasticity profiles for each consumer over different hours for different prices.

In Figure 1 and Figure 2, the estimate 'p' is represented as "Response Probability". The plot in Figure 1 shows the p for client D0767 data including all three prices for all 24 hours and Figure 2 shows the mean values of the responses for each price signified by distinct colors and their relative probability of response for each hour. This is represented as:

$$\sum_{i=1}^{k} c_{i,h}^{k} = n_h \qquad (9)$$

where, $k$ = 3 prices; $c$ = observed elasticity; $n$ = (sum of weights) $_h$

where consumer is kept constant to one client, and "h" and "n" for all price level (Default, High and Low).

### 3.2 Numerical Results

The Dirichlet-distribution is initialized to a 1s vector for each price for every hour, making it a 3x24 matrix uniform prior. A separate Dirichlet-distribution (theta (θ)) for each hour is created where the sum of probabilities across prices is 1 for each hour. We sample this multinomial with our observed data, to identify the posterior probability distribution of "c" for these conditions. In our example shown in Figure 2, k=D0767 for Price = all three tariffs and their corresponding n (sum of the weights) for each hour. Then Figure 3 and Figure 4 display the results obtained for consumer D0323 and Figure 5 and Figure 6 present the results for consumer D0644.

One interesting behavior to note for consumer D0323 is that the response probability (p) is much wider compared to D0767. This is evident looking at Figures 4 where consumer D0323 is susceptible to high prices and thus shows varied probability in accepting to participate in DR events compared to lower prices which has the exact same behavior as that for default prices. This leads us to infer that D0323 is a more reliable consumer for low prices than high prices. In this case, a more reliable consumer refers to the expected action out of a consumer for a given intervention signal. As seen from Figure 4, consumer D0323 is readily available to change their consumption behavior (i.e., increase their consumption) when experiencing a low-price tariff, compared to their willingness to not alter their consumption (i.e., reduce consumption) when experiencing a high-price tariff.

On the other hand, consumer D0644 as seen from Figure 5 and 6 exhibits a behavior that is contrary to the one of consumer D0323 where they tend to be more reliable for high prices following behaviors similar to default prices. The periods between 10 h and 17 h display a drop in the response probability for low prices. This drop is likely due to the lack of consumption activity in that period, where lowering the price is not incentive enough to increase consumption.

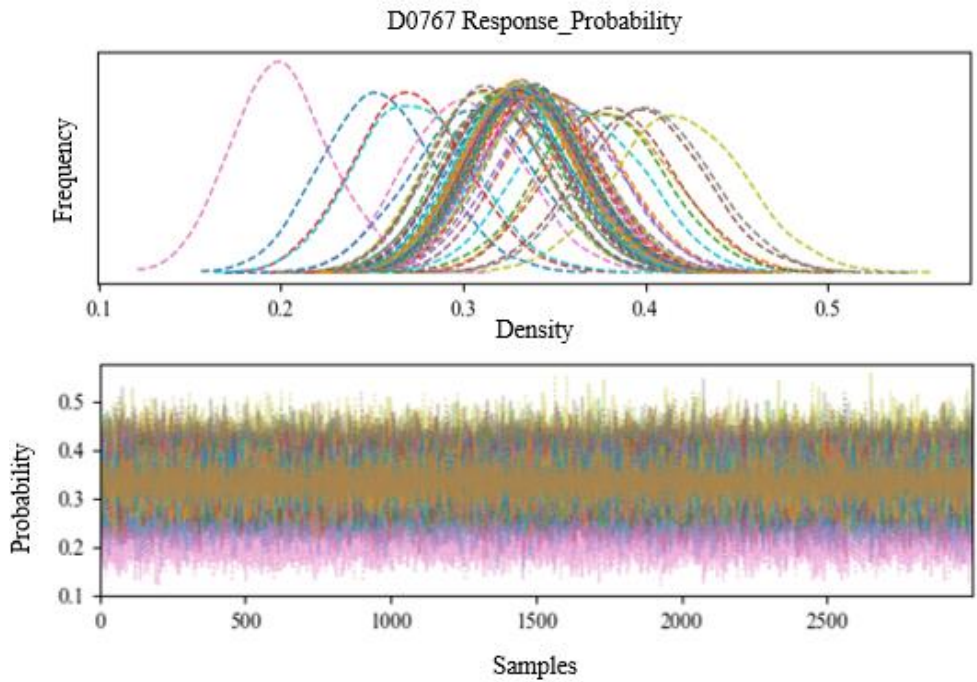

*Figure 1 Estimated probability of response for a client D0767 for all three prices under 24 hours. Based on Dirichlet-Model computed for 3000 samples.*

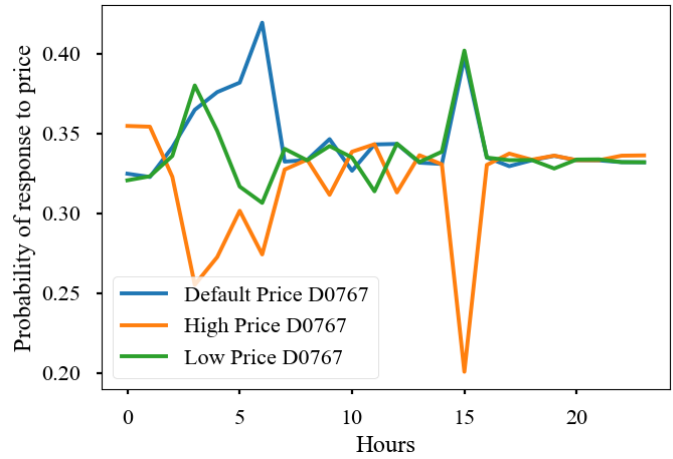

*Figure 2 Consumer D0767's estimated probability of response to all three prices for each hour from the Dirichlet-Model*

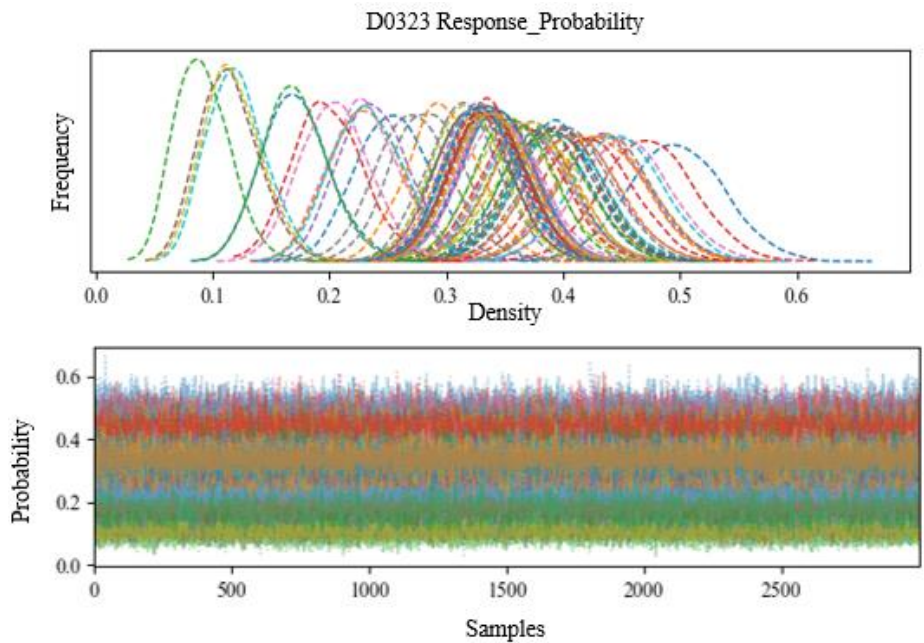

*Figure 3 Estimated probability of response for a client D0323 for all three prices under 24 hours. Based on Dirichlet-Model computed for 3000 samples.*

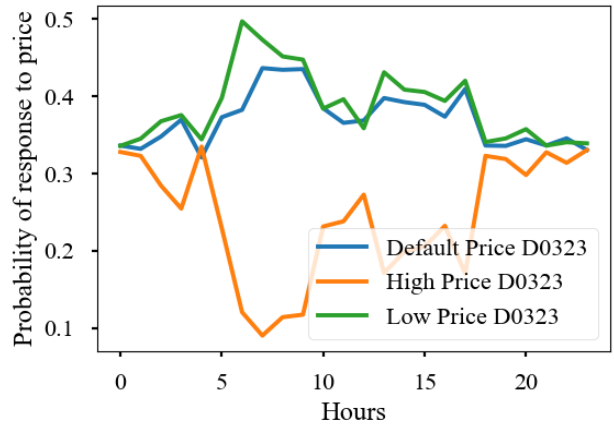

*Figure 4 Consumer D0323's estimated probability of response to all three prices for each hour from the Dirichlet-Model*

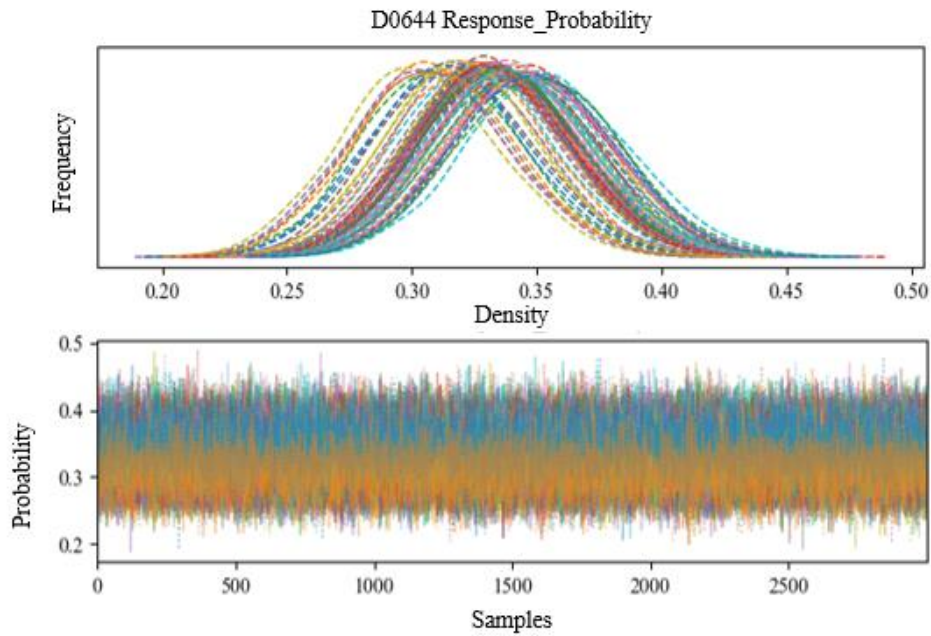

*Figure 5 Estimated probability of response for a client D0644 for all three prices under 24 hours. Based on Dirichlet-Model computed for 3000 samples.*

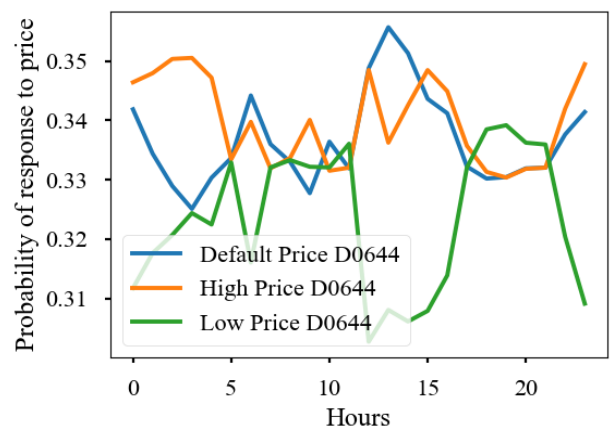

*Figure 6 Consumer D0644's estimated probability of response to all three prices for each hour from the Dirichlet-Model*

Comparing the estimated response probability plots for the three clients shown above, it is evident that some clients are more susceptible to price changes and show elasticity based on the prices they experience. This is seen in client D0767 and the price effect is pronounced for client D0323. On the other hand, client D0644 seems not to be very price-sensitive, but it still shows probabilities of higher elasticity for different hours. This is clearly seen in the probability density plot of Figure 5 and Figure 6. It is also imperative for us to identify the difference in response probability for a pool of consumers. This characterizes how the consumers would react when they are pooled into a selected group for a given price. Thus, our data fed into the Dirichlet-model is modified to replicate a selected group of consumers for a single DR price they have experienced. Figure 7 depicts the probability of response for 5 clients grouped under the Low DR price. This is represented as:

$$\sum_{i=1}^{k} c_{i,h}^{k} = n_h \quad (10)$$

Where, $k$ = D0767, D0323, D0644, D0806 and D0556

$n$ = *(sum of weights)* $_h$

In our example shown in Figure 7, k = D0767, D0323, D0644, D0806 and D0556, with each price level kept constant (either Default, High or Low) and their corresponding n (sum of the weights) for each hour. The Dirichlet-distribution is initialized to a 1s vector for each consumer for every hour for a fixed price, making it a 5x24 matrix with uniform prior. A separate Dirichlet-distribution (theta ($\theta$)) for each hour is created where the sum of probabilities across prices, 1 for each hour. We sample this multinomial with our observed data, to identify the posterior probability distribution of "c" for these conditions.

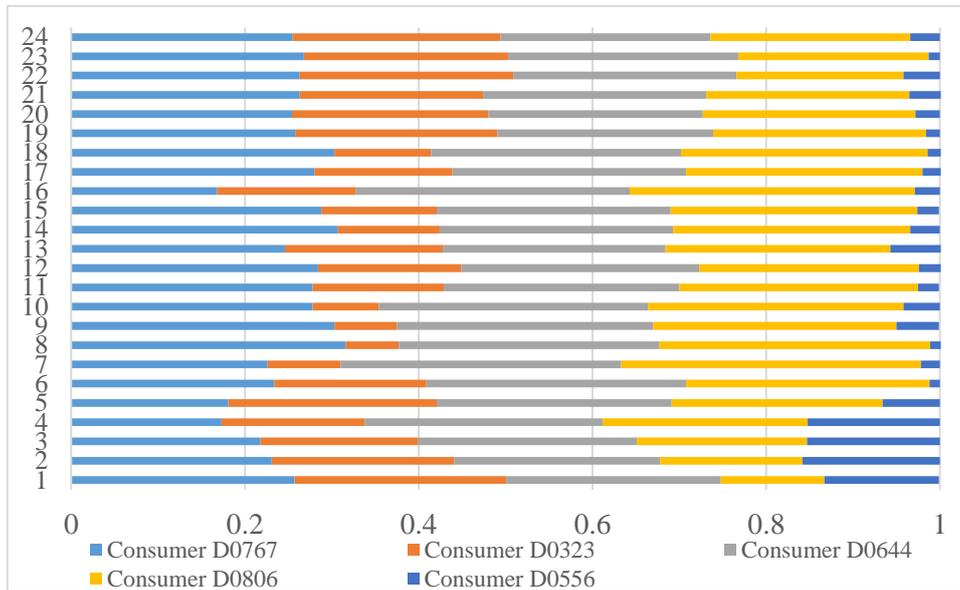

*Figure 7 Probability of accepting or acting to change consumption for a given price signal. Shows Consumer Probability score for 5 consumers under Low Price Tariff.*

## 4. Elasticity Behavior Model

*4.1 Methodology*

The second original contribution for this framework is based on a methodology using tailor-made (bespoke) Generalized Linear Model (GLM) for each consumer. To understand GLM with Bayesian Inference, we briefly describe the Bayesian linear regression. As mentioned earlier, for inference studies using probabilistic models, we deal with distributions, so it is necessary to specify the linear regression as a distribution. A general standard linear regression is given by Equation (11a) and its relative probabilistic reformulation is given by Equation (11b), a multiple-linear regression is given by Equation (12a) and its corresponding probabilistic reformulation is given by (12b).

We note that the outcomes, y and Y, are normally distributed with their respective means µ and Xβ along with a standard deviation of σ. These variables (priors), that define the distribution of our outcome, are also considered to be obtained from some distribution. We also do not get a single estimate for $\alpha$ and $\beta$, but instead a density distribution about how likely different the values of β are. This quantifies the uncertainty in the estimation, and how spread out they are in identifying y (or Y).

$$\mu = \alpha + \beta_1 X_1 + \beta_2 X_2 \tag{11a}$$

$$y \sim \mathcal{N}(\mu, \sigma^2) \tag{11b}$$

$$f(X) = \beta_0 + \sum_{j=1}^{p} X_j \beta_j + \varepsilon \tag{12a}$$

$$Y \sim \mathcal{N}(X\beta, \sigma^2) \tag{12b}$$

$$\text{with } \varepsilon = \mathcal{N}(0, \sigma_\varepsilon^2), \beta = \mathcal{N}(0, \sigma_\beta^2), \alpha = \mathcal{N}(0, \sigma_\alpha^2) \tag{13}$$

Our goal in this model is to identify how the consumers react to different prices based on the ones they have already experienced. This is one of the reasons why we chose to use probabilistic techniques to address this price responsive behavior with uncertainties to quantify our model, as one cannot accurately identify the exact behavior of a consumer regarding prices they have not experienced (unobserved data). Thus, instead of relying on classical multiple-linear regression, we use a probabilistic approach. The data for the elasticity behavior model is derived from the available features from the consumption data, such as their temperature profiles, time components and data derived from the causality framework – 24 hour average consumption along the year, the actual elasticity, the prices experienced (Default, High and Low for LCL dataset). The temperature component is split into temperature_high (th), temperature_low (tl) and temperature_average (ta) for each hour and each price level. The model also includes a "difference in average" component which corresponds to the difference between the default price consumption average to the high-price and low-price consumption averages. This component is necessary to define the change in elasticity the consumer shows for the change in price.

*Model Specification:* the model to be developed encompasses our understanding of how the elasticity is influenced by the data we have. Using this model, we want to identify what are the values of the model's parameters that best explain the observed elasticity. As mentioned, we carefully identify the priors for each consumer, and form the stochastic variables. Table 1 describes the Elasticity Behavior Model coefficients and their corresponding distribution type that will influence the dependent stochastic variable, with each of these coefficients dependent on the hour and price for each sample datapoint.

| 1 | **Component Identifiers** | | |
|---|---|---|---|
| 1.1 | Time Component (Hour) | $h$ | 1-24 |
| 1.2 | Price | $p$ | Default = 0.1176 \| High = 0.6720 \| Low = 0.0399 |
| 2 | **Model Coefficients** | | |
| 2.1 | Intercept | $\beta_{0,h}$ | Normal Distribution |
| 2.2 | Price | $\beta_{1,h}$ | Normal Distribution |
| 2.3 | Temperature High | $th_{p,h}$ | Normal Distribution |
| 2.4 | Temperature Low | $tl_{p,h}$ | Normal Distribution |

| 2.5 | Temperature Average | $ta_{p,h}$ | Normal Distribution |
| 2.6 | Average Consumption | $yavg_p$ | Normal Distribution |
| 2.7 | Difference in consumption | $ydiff_p$ | Normal Distribution |
| 2.8 | Observed Elasticity likelihood (dependent variable) | $y$ | Student-T distribution |
| 2.9 | Degrees of freedom | $\nu$ | Uniform distribution |
| 2.10 | Standard Deviation | $\sigma$ | Exponential distribution |

*Table 1. Elasticity Behaviour Model Components*

The Student-T prior for the observed elasticity (dependent variable) y is considered instead of a normal distribution, because it is better at handling outliers as it has a longer tail. From Figures 8, and 9, we can see that our elasticity data has a significant amount of datapoints in the tail end of the distribution, and it is pertinent to understand the elastic behavior of the consumers under all conditions. Using the student-t distribution indicates our model to incorporate these outliers also as data points to consider as they are not indicators of data abnormality. The Student's T log-likelihood of a normal variable (x) with gamma (Γ) distributed precision is given by (14), where the scalar variable μ denotes the mean of the stochastic variable from our model (elasticity), ν degree of freedom with a uniform distribution between 0 and 1, and λ given by the standard deviation (σ) with an exponential log-likelihood.

$$y = f(x \mid \mu, \lambda, \nu) = \frac{\Gamma\left(\frac{\nu+1}{2}\right)}{\Gamma\left(\frac{\nu}{2}\right)} \left(\frac{\lambda}{\pi\nu}\right)^{1/2} \left[1 + \frac{\lambda(x-\mu)^2}{\nu}\right]^{-\frac{\nu+1}{2}} \quad (14)$$

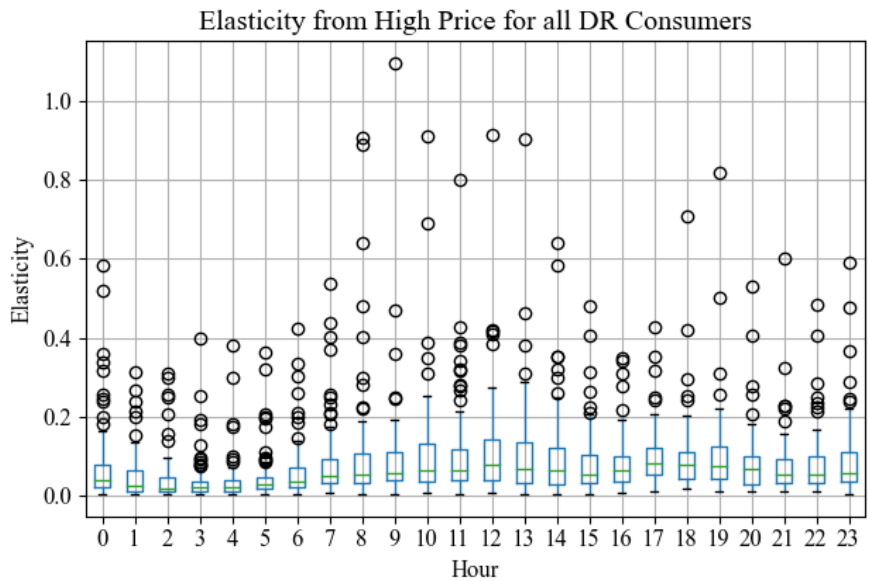

*Figure 8 Grouping of elasticity values for all consumers for each hour under the high DR price*

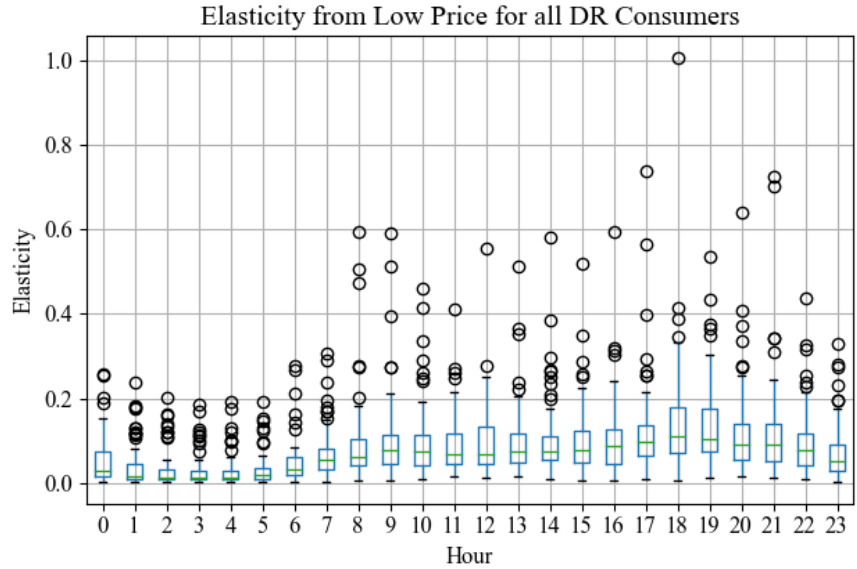

*Figure 9 Grouping of elasticity values for all consumers for each hour under the low DR price*

The mean $\mu$ (stochastic variable) of Equation (14) is the main component of our elasticity model that describes the prior distributions over the observed data to link them with the response variable $y$. We also are specifically interested in whether different hours actually have different relationships (slope) and different intercepts. This way, we can have the aggregated mean ($\mu$) for all hours, as well as individual hours, for every price along the distribution. Using Equation (15), we estimate the probability distribution of $\beta_0$ and $\beta_1$ for each hour (note that all components are indexed for each price). We also use a $price^2$ and $price^3$ terms to help with the model convergence.

$$\mu = \beta_{0,h} + \beta_{1,h}(price)^2 + \beta_{1,h}(price)^3 + th_{p,h}(Temp\_high) + tl_{p,h}(Temp\_low) + ta_{p,h}(Temp\_avg) + yavg_p(consumption\_avg * hour) + ydiff_p(consumption\_difference * hour) \qquad (15)$$

## 4.2 Numerical Results

Similar to the Dirichlet-model, we implemented this Elasticity Behavior model using MCMC in PYMC for 2000 samples (increasing the number of samples increases computation time, 2000 was chosen as we had convergence already). Our model draws 2000 samples from the posterior in parallel computing (2000 * 4 for a 4-core machine). We adjusted the model to use the first set of 1000 samples as tuning samples to learn from, and the inferences are made only after the first set of 1000 samples are used.

Model data contains the components responsible to make the desirable outcome $y$ for each hour and each price, making it a dataset with 72 outcomes for every consumer. The posterior plot for consumer D0346's model components is given in Figures 10, 11 and 12. They describe the marginal posterior distribution for the stochastic variables for each hour running from b_0 0 to b_0 23 for coefficient $\beta_0$ stacked on top of each other and b_1 0 to b_1 23 for coefficient $\beta_1$ also stacked on top of each other. These Figures summarize the posterior distributions of the parameters (from the provided prior distributions and hyper-parameters) and present a 95% credible interval and the posterior mean. The plots below are constructed with the 2000 samples from each of the 4-cores (chains), pooled together. By looking at Figure 11 and 12, we also see that the marginals for $\beta_0$ and $\beta_1$ have distinctive differences in their values between hours, compared to the other model components shown in Figure 10. This is the specific reason why we chose to incorporate hourly changes to the intercept and price in our model. The more the measurements per hour (more than 72 outcomes), the higher will be our confidence in estimating the posterior density distribution. Looking at these distributions, we can gather information on how representative we think our estimate is for the given model.

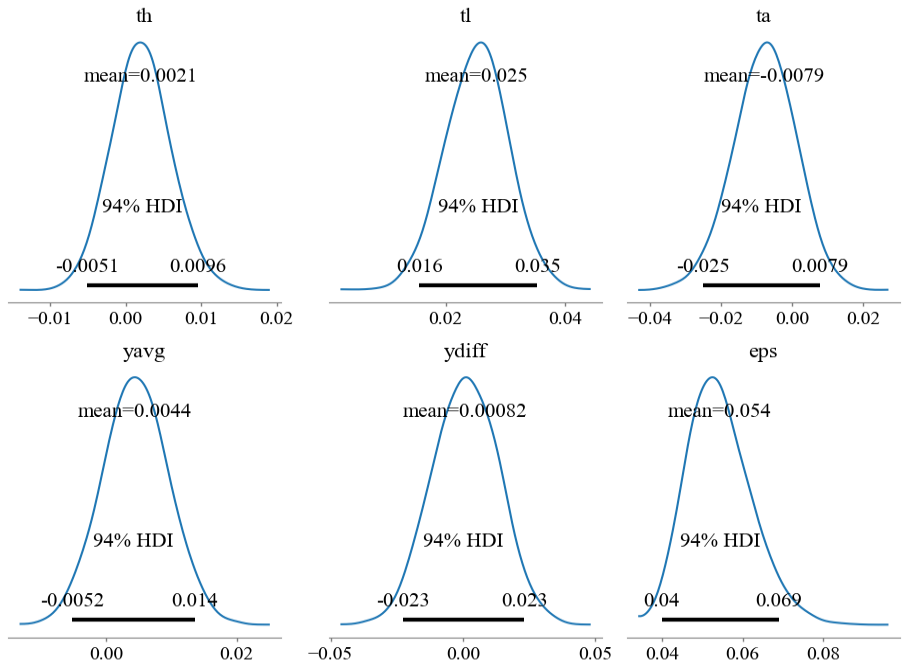

*Figure 10 Marginal posterior distribution for stochastic variables $th_p, tl_p, ta_p, yavg_p, ydiff_p$ and $\sigma$ from the elasticity behavior model sampled based on their user specified prior distributions for consumer D0346*

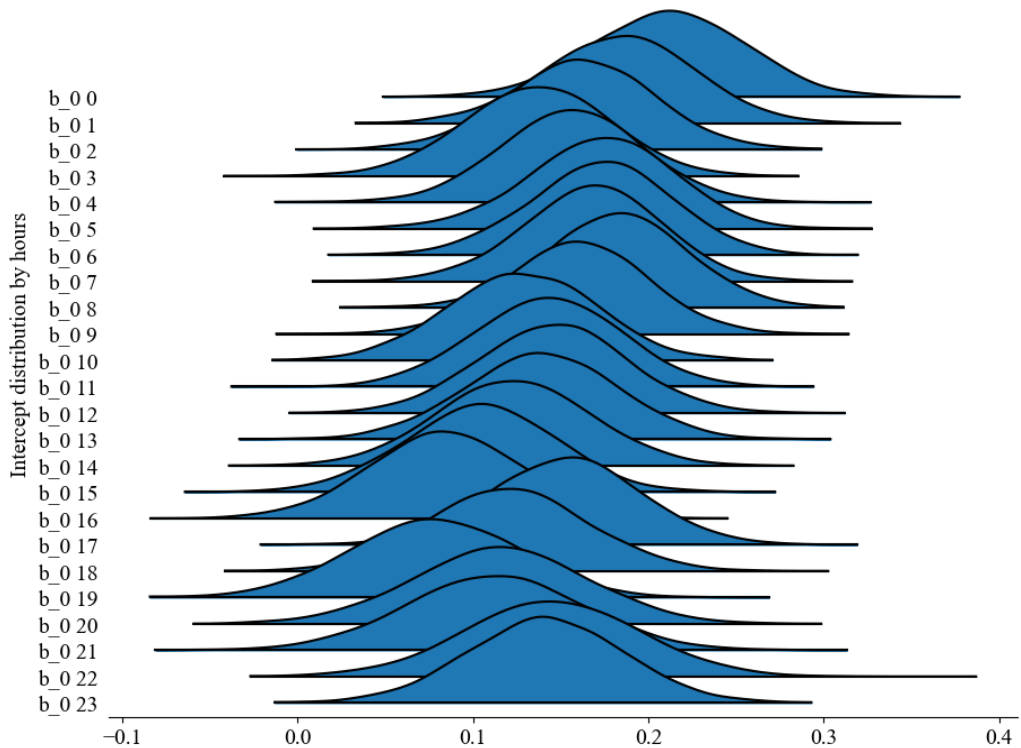

*Figure 11 Marginal posterior distribution for stochastic variable $\beta_{0,h}$ and $\sigma$ from the elasticity behavior model sampled based on their user specified prior distributions for consumer D0346*

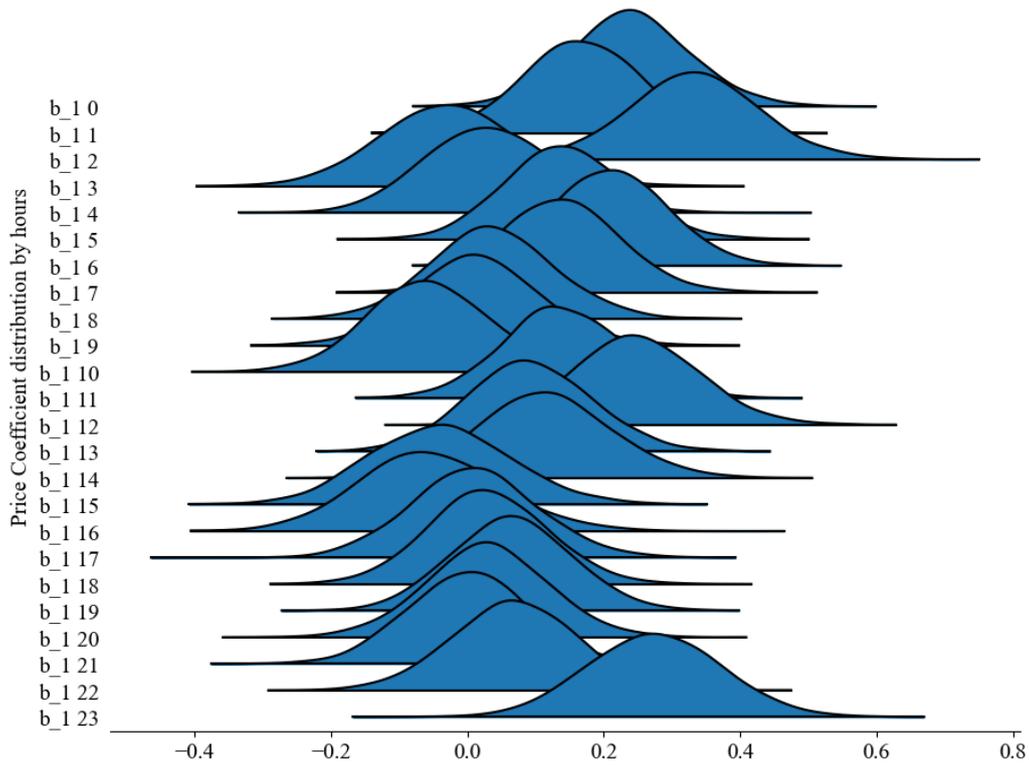

*Figure 12 Marginal posterior distribution for stochastic variable $\beta_{1,h}$ and $\sigma$ from the elasticity behavior model sampled based on their user specified prior distributions for consumer D0346*

From the obtained traces containing the family of posteriors sampled for each consumer, conditioned on the observations, we can plot the regression lines to compare them with their mean and uncertainty (95% elasticity quantile). We plot the regression lines for some consumers in Figure 13 that show the estimated elasticity behavior measurements for different prices.

At this point, we must understand that our model derived these distributions based on 3 prices that the consumers experienced during the recorded LCL DR program. In case we have a scenario where consumers have experienced many price signals, for example dynamic DR tariffs, then this model can be replaced with a comprehensive behavior model to get a better fit and estimate. This is true as we will have more data points for identifying better behavior pattern.

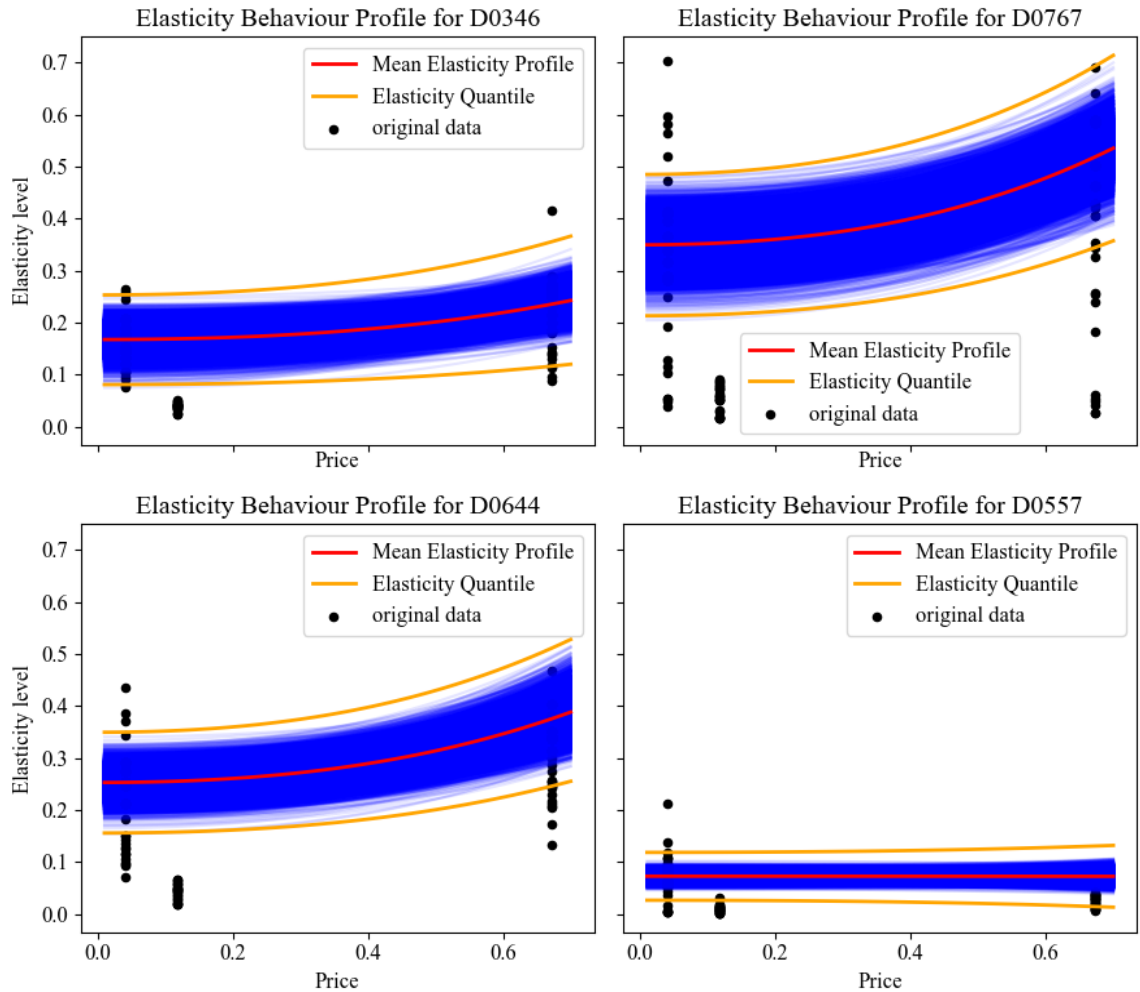

*Figure 13 Sampled range of posterior regression lines from the elasticity behaviour model for consumers D0346, D0767, D0644 and D0557*

These regression lines are formed from the posterior predictive distribution recreating the data based on the parameters found at different moments in the chain. The recreated or predicted values are subsequently compared to the real data points. The thick red line represents the mean estimate of the regression line of the individual consumer and the thinner blue lines are the individual samples from the posterior that give us a perspective of uncertainty in our estimates. The elasticity behavior model can be used to identify how a consumer reacts to different prices for a given time-period and temperature conditions.

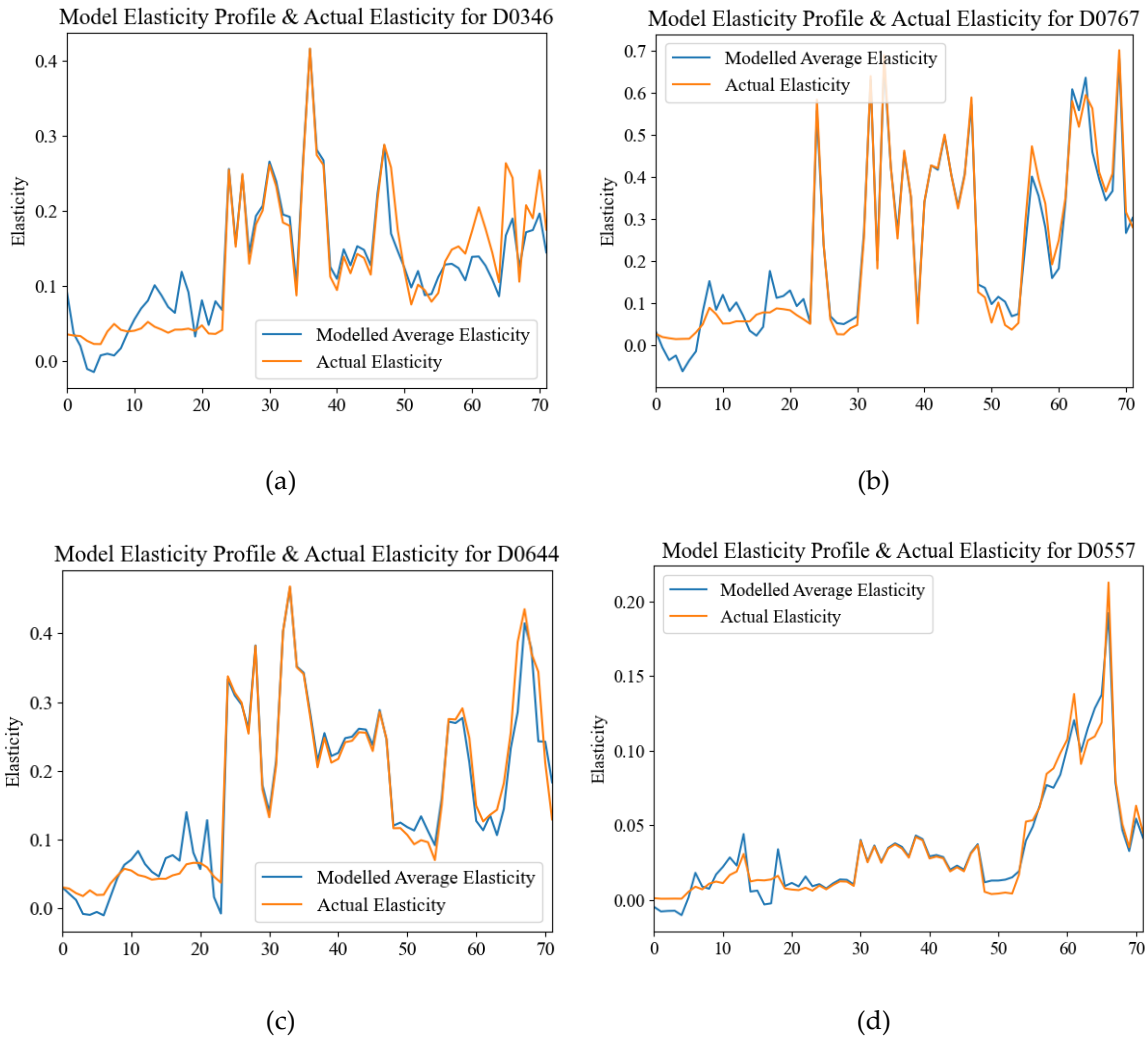

*Figure 14 Average elasticity profile compared to the actual elasticity profile from the causality model, for all 72 response variables (y) for consumers (a) D0346, (b) D0767, (c) D0644 and (d) D0557*

Figure 14 compares our elasticity behavior model's mean estimates with the actual elasticity provided by the causality model for each 72 response variables and depicts our model convergence. The 1st 24 response variables (0-23) correspond to the default DR price - 0.1176 £/kWh, response variables 24 to 47 correspond to high DR price - 0.6720 £/kWh, and response variables 48 to 71 correspond to low DR price - 0.0390 £/kWh. It should be noted that our model assumes that all the prices considered in these posterior estimates are new DR prices, and this is the reason we see smaller deviations in our modelled average elasticity in the first 24 response variables (which correspond to the default DR price). The posterior probability distribution for each of the 72 response variables (*y*) for our estimates of elasticity with the elasticity behavior model is shown in Figure 15 for consumer D0346.

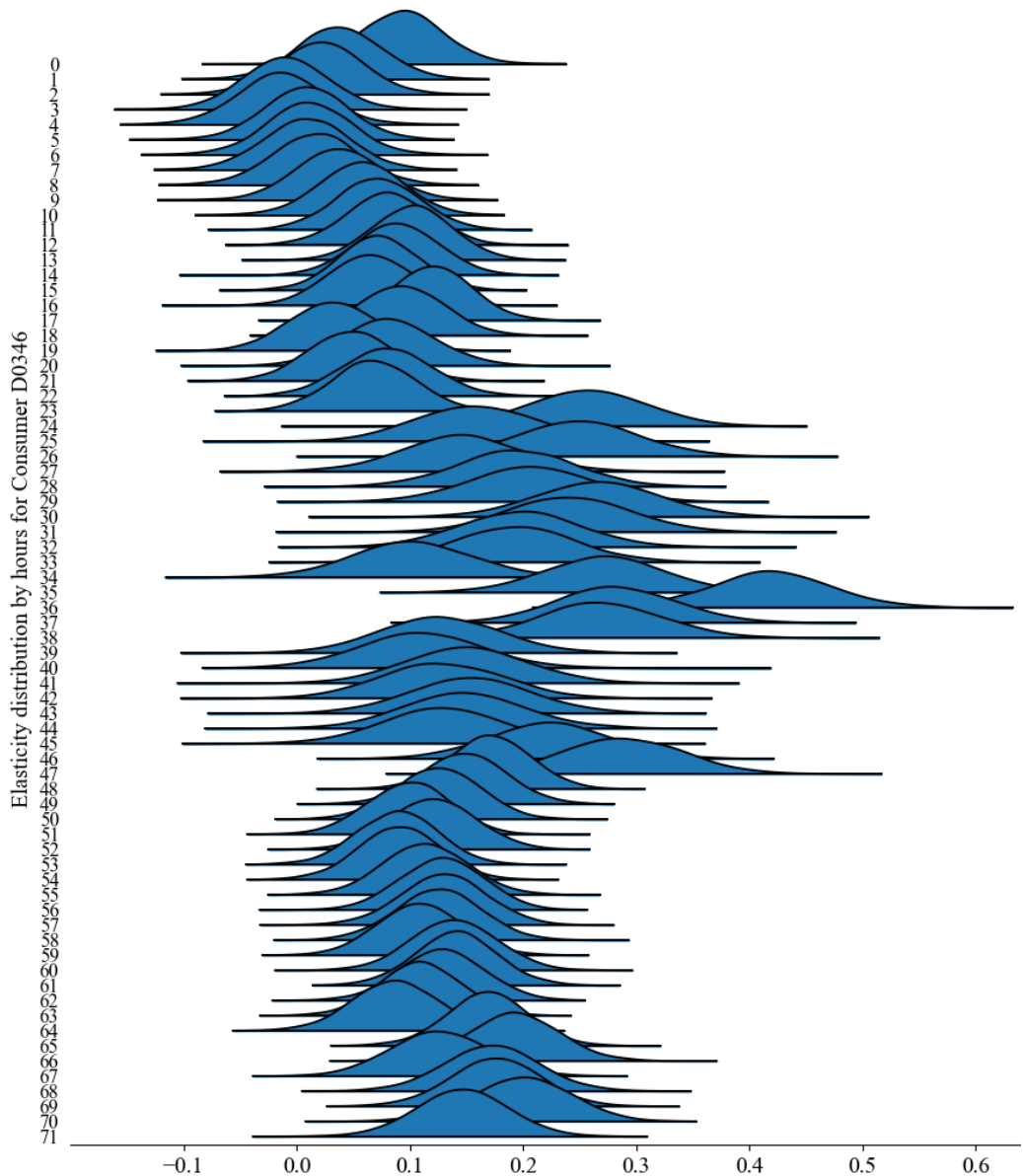

*Figure 15 Posterior probability density distribution of elasticity for each of the 72 response variables obtained by the elasticity behaviour model*

The model was constructed to identify the effects from the marginals for $\beta_0$ and $\beta_1$ as they have distinctive differences in their values between hours of the day (Figure 11 and 12). From our causality framework, it is also evident that the consumers behave differently for different hours of the day. To show this pattern in our estimated behavioral model, we analyzed the impact of price change on elasticity for each hour as shown in Figures 16 and 17. This gives us an inference that consumer D0346 shows a typical elasticity profile (expected by utilities or retailers) where the consumer is quite price responsive the early hours of the day. During the rest of the hours, namely on midday and late-night hours, this consumer's behavior is quite inelastic.

Compared to this, consumer D0767 seems to have a very price responsive behavior throughout the day with one of the most elastic periods being hour 10:00 am of the day. Typically, this cannot be seen in a general price responsive model, where they will follow the elasticity profile shown by consumer D0346 as a general price responsive behavior for all their clients. This elasticity behavior model is constructed in such a way that the utility or retailer can tune the features enumerated in Table 1 to accurately identify the elasticity behavior profile for each consumer and use the consumer probability score model to group and identify the exact probability of consumer responsiveness to the new DR price offers. This way, the utility or retailers can know how a new DR strategy can perform amongst its clients with greater reliability.

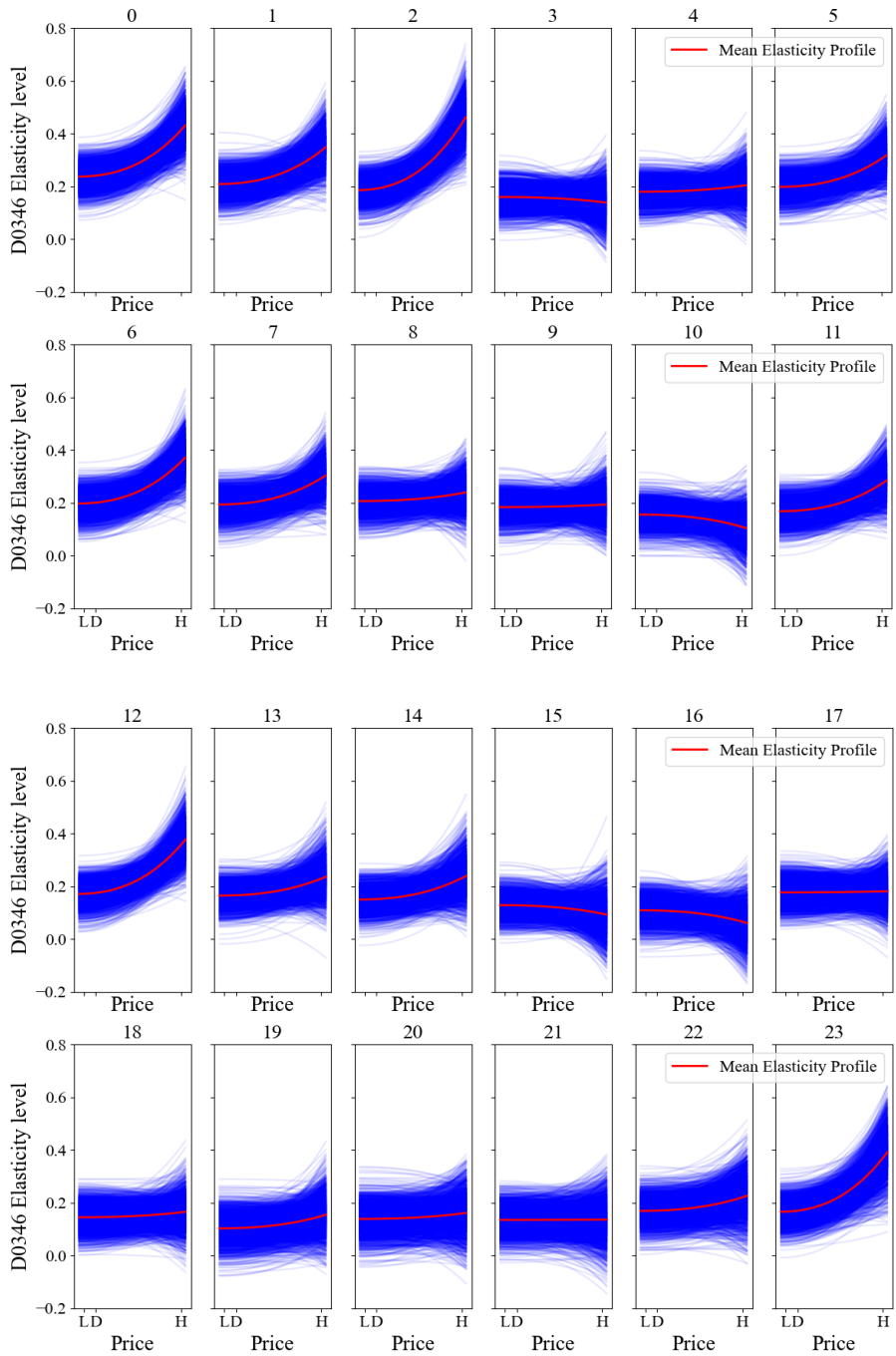

*Figure 16 Sampled posterior regression lines from the elasticity behaviour model for consumers D0346 for each hour*

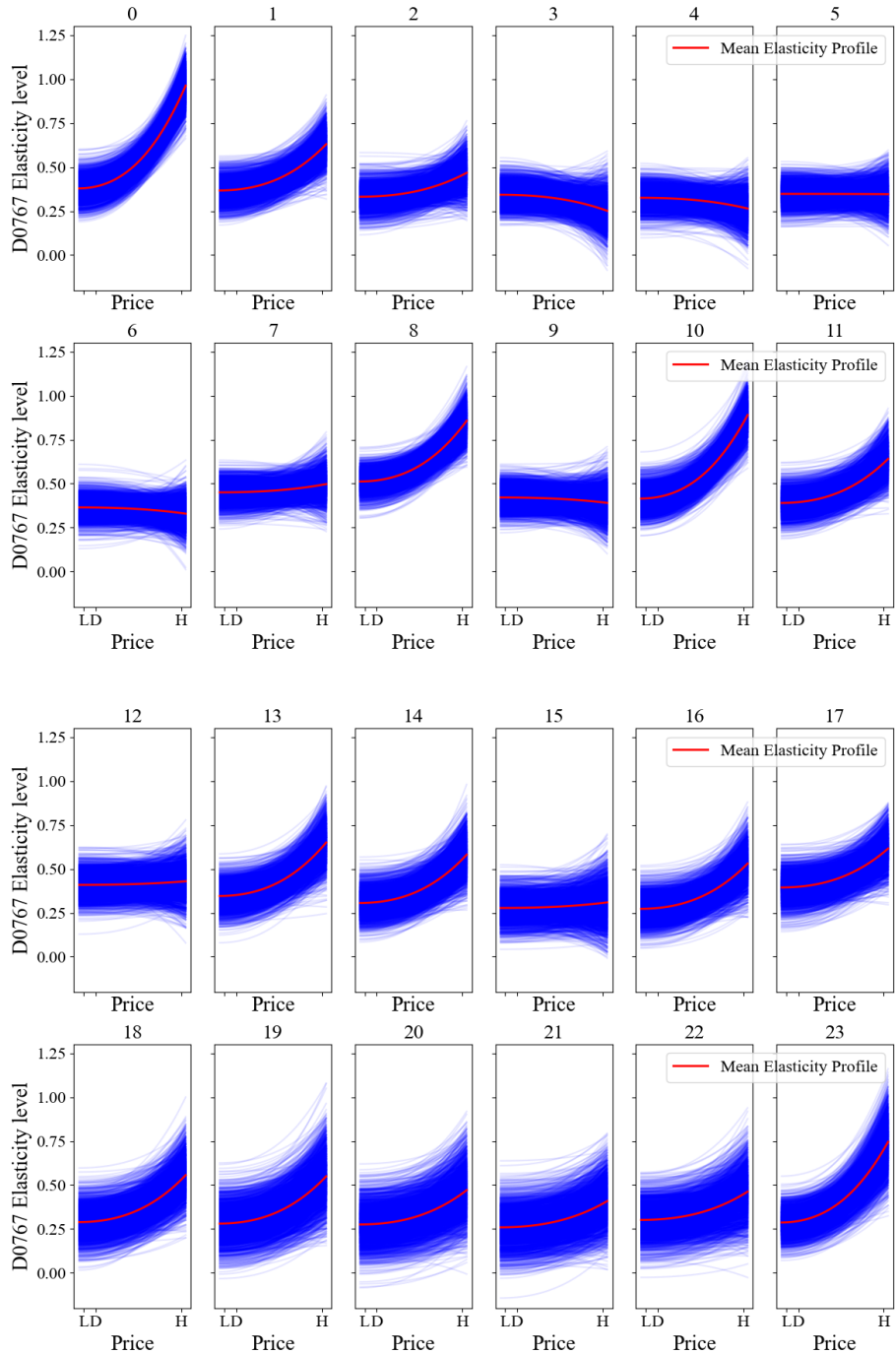

*Figure 17 Sampled posterior regression lines from the elasticity behaviour model for consumers D0767 for each hour*

**5. Conclusions**

As more and more retailers and utilities turn their focus towards residential DR programs, they emphasize the urgency to ensure reliability from their participants to make these programs more viable. The methodology described in this paper tackles this issue by identifying the consumers reliability of provision of demand response and also a way to measure how and for whom the DR strategy works.

The first model, Consumer Probability Score model used Dirichlet multinomial distribution to address the user engagement question and accurately quantifies the likelihood with which the consumer reacts to DR signals. We were able to make a fair assessment of the consumer's willingness to participate in DR events. Using the LCL dataset, we conclude that consumers who show low variations in response probability are more reliable and less risky to be considered for targeted DR signals rather than consumers who have high variations in their responses. Regarding the second model, the Elasticity Behavior Model uses a Bayesian probabilistic approach and identifies the actual load adjustment an individual contracted DR client provides for different price levels they can experience. This allows the retailers to understand the full potential of consumption elasticity behavior from their contracted clients from the already available data without the need to send new price signals and strategies to the same consumers. Thus, this work provides a holistic approach to capture the customers uncertainty in modeling the consumer selection problem for new targeted DR strategies.

The framework proposed in this paper was developed with the goal for it to be scalable. This framework can be effectively extended to other consumers such as small and medium service buildings and even community buildings where extracting a larger amount of controlled and aggregated services can become highly effective. With predicting a consumer's kWh consumption becoming a more widely and common study, using the elasticity measure as a feature for such model would help the model in achieving higher accuracies. This has already been proved by a research work [43, 44] where the authors, using Conditional Variational Autoencoders, generate daily consumption profiles of consumer segmented in different clusters (based on response to electricity tariff). We used dynamic Time-Of-Use (TOU) tariffs for our elasticity behavior model. A continuous dynamic pricing such as RTP tariffs can bring more granularity and better insights into a consumer's consumption behavior. Extending this work towards this direction would enable understanding what is the best-fit pricing strategy for residential consumers before hitting elasticity saturation and eventual indifference towards price changes.

**Appendix A**

**6. Causality Inference Algorithms**

Data analysis and causal inference for our framework was performed using Robin's g-method to estimate the average consumption and elasticity for each hour. The Robin g-method enables the identification and estimation of the effects of generalized treatment, exposure, and intervention plans. It provides consistent estimates of contrasts of average potential outcomes under a less restrictive set of identification conditions than standard regression models [45]. The estimated consumption elasticity is then pooled together with all clients and ranked by hours. The modeled approach helps in making a fair estimate of whether the consumers introduced any changes to their consumption based on DR signals, as it can differentiate between a normal consumption and a DR stimulated consumption. The algorithm used in our consumer elasticity model uses the parametric g-formula (as the causal effect estimator) and arbitrary machine learning estimators to analyze and plot causal effects [46]. Causal effect refers to the distribution or conditional expectation of Y given X, controlling for an admissible set of covariates, Z, to make the effect identifiable. Covariates Z are exogenous variables whose value is determined outside the model and is imposed on the developed model. In our analysis (simple depiction in Figure A.1), X, being price, and Z, being time, are indeed correlated; more precisely, they will be statistically dependent.

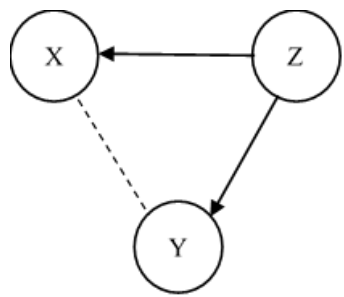

**Figure A.1** Causal Graph of X and Y related by a common cause

This is basically a fork graph showing that X and Y are related with a common cause, Z. Variables X and Y can be considered independent, but are conditionally dependent, as one can trace the variables upstream from it and conditioning on this common effect will make upstream factors dependent. To validate our assumptions, we ran our collected dataset (as mentioned in Chapter 3) with Tetrad developed by the Center for Causal Discovery (CCD) [47]. Using their Fast Greedy Equivalence Search (FGES) Algorithm for our continuous (consumption) variables, we were able to hypothesis causal relationships between consumer's consumption (kW), price, day, hour of the day and temperature as show in Figure A.1.

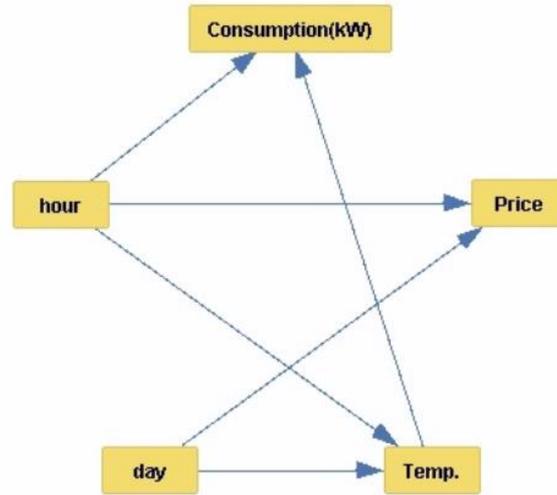

**Figure A.2** Hypothesised Causal relationship between collected data variables

By inspecting this figure, we can infer that consumption is clearly dependent on "hour" and "temperature" variables, while price is directly dependent on "hour" and "day" variables. It should be noted that the graph in Figure A.1 can be expanded further, if we provide supporting information for how the "prices" were formed, external factors affecting the "temperature", and other notable consumer specific information such as the occupancy rate, household building information etc. However, these items are not covered in this research, as our goal is to identify consumer's flexibility based on non-invasive information collected from smart meters.

It is the main objective of this work to recover unbiased estimates of causal effects between intervention (price) – outcome (consumption) relationship. It should be noted that it is necessary for variable Z to satisfy the back-door criterion, meaning that it should block all backdoor paths from treatment to the outcome variable and it should not include any descendants of the treatment variable [48]. If Z satisfies this back-door criterion, then the probability of change in Y because of X given Z is indicated by Equation (A.1). The operator do() marks an action or an intervention in the model, which in our case is the DR event price (X). As shown in Equation (A.2) the causal conditioning $P(Y|do(X=x))$ is the probability of change in Y because of the intervention X=x with summation over all the values of Z. The model used to minimize the mean-squared error and do the controlling is a Kernel regression model, which gives us the conditional expectation of our random variable (Y) as shown in Equation (A.3a). The objective is to find a non-linear relationship between the random variable (Y) and the price (X) by controlling on exogeneous variables (Z).

$$P(Y, X | Z = z) = \frac{P(X, Z, Y)}{P(Z)} = P(Y|Z)P(X|Z) \tag{A.1}$$

$$P(Y|do(X = x)) = \sum_Z P(Y|X, Z)P(Z) \qquad (A.2)$$

$$E(Y|x) = \sum_Y YP(Y|x) = \sum_{Y,Z} YP(Y|X, Z)P(Z) \qquad (A.3a)$$

$$m(x) = E[Y|do(X = x)] = \sum_Z E[Y|X, Z]P(Z) \qquad (A.3b)$$

$$E[Y|(X_j)] = \sum_{h=1}^{24} \frac{1}{N} \sum_{i=1}^{N} E[Y|X_j, (Z_{hi}, Z_{2i}, Z_{3i}, Z_{4i})] \qquad (A.4)$$

$$Z - \text{Interval} = E[Y|X, Z] \pm z_{\alpha/2}(\sigma) \qquad (A.5)$$

The model used by default to minimize the mean-squared error and do the controlling is a Kernel regression model which gives us the conditional expectation as shown in Equation (A.3b). The causal dataframe helps in finding the true dependency between the X (price) and Y (consumption) with the confounders as the calendar variables, we consider X: price, Y: consumption and Z: calendar variables (hour - Z1, day - Z2 and week - Z3). For every hour (Z1), the causal effect of (X) on (Y) is obtained for each DR price ($X_j$) with constant (Z1) and variable (Z2 & Z3) with i=1 to N running over all data points as shown in Equation (A.4). These calendar variables are confounders that have effects on both price and consumption. The causal analysis model works by controlling the Z variables when trying to estimate the effect of variable X on a continuous variable Y (Equation A.4). The model returns Y estimates (E[y]) at each X=x value for every Z and provides the upper and lower average consumption limits of the Y variable, which is used to determine the average elasticity. The upper and the lower consumption limits of the average Y variable (for each price) is the z-intervals calculated using the confidence levels (CI = 95% for our model). The z-interval for our consumption average Y estimates is shown in Equation (A.5) where $z_{\alpha/2}$ is the alpha level's z-score for a two tailed test (based on the value of CI) and σ is the standard deviation of our average Y estimate. The constructed causality model is then used to provide the estimates for each consumer demand on each hour for each price.

**Funding:** This work was financially supported by the European Union's Horizon 2020 research and innovation programme, under Grant agreement No 857237 (InterConnect—Interoperable Solutions Connecting Smart Homes, Buildings and Grid). Kamalanathan Ganesan was also supported by the European Social Fund (FSE) through NORTE 2020 under PhD Grant NORTE-08-5369-FSE-000043. The sole responsibility for the content lies with the authors. It does not necessarily reflect the opinion of the CNECT or the European Commission (EC). CNECT or the EC are not responsible for any use that may be made of the information contained therein.

**Conflicts of Interest:** The authors declare no conflict of interest. The funders had no role in the design of the study; in the collection, analyses, or interpretation of data; in the writing of the manuscript, or in the decision to publish the results.